# Foundations of a Knee Joint Digital Twin from qMRI Biomarkers for Osteoarthritis and Knee Replacement

Gabrielle Hoyer[1,2,3,*], Kenneth T Gao[1,2,3], Felix G Gassert[1], Johanna Luitjens[1], Fei Jiang[1], Sharmila Majumdar[1,2,3], Valentina Pedoia[1]

[1]Department of Radiology and Biomedical Imaging, University of California, San Francisco, San Francisco, CA, USA
[2]Department of Bioengineering, University of California Berkeley, Berkeley, CA, USA
[3]Department of Bioengineering, University of California San Francisco, San Francisco, CA, USA

* Correspondence: gabbie.hoyer@ucsf.edu

## Abstract

This study forms the basis of a digital twin system of the knee joint, using advanced quantitative MRI (qMRI) and machine learning to advance precision health in osteoarthritis (OA) management and knee replacement (KR) prediction. We combined deep learning-based segmentation of knee joint structures with dimensionality reduction to create an embedded feature space of imaging biomarkers. Through cross-sectional cohort analysis and statistical modeling, we identified specific biomarkers, including variations in cartilage thickness and medial meniscus shape, that are significantly associated with OA incidence and KR outcomes. Integrating these findings into a comprehensive framework represents a considerable step toward personalized knee-joint digital twins, which could enhance therapeutic strategies and inform clinical decision-making in rheumatological care. This versatile and reliable infrastructure has the potential to be extended to broader clinical applications in precision health.

## Introduction

Osteoarthritis (OA), a multifaceted degenerative joint disease, substantially contributes to chronic pain and disability in the United States, affecting an estimated 30.8 million adults.[1] With the growing prevalence of risk factors such as obesity, depression, and aging, the incidence of OA is expected to rise substantially in the coming years, placing an even greater burden on public health systems.[2–4] However, OA is not solely a concern for the elderly; younger individuals, particularly those with joint injuries, are increasingly at risk, leading to a surge in knee replacement surgeries.[5,6] The societal impact of OA is profound, encompassing reduced quality of life, increased healthcare costs, and lower employment rates.[7–10]

Even so, the management of osteoarthritis remains hindered by the scarcity of advanced prognostic tools, preventive therapies, and non-invasive treatments. While imaging methods such as radiography and conventional MRI do well to detect structural damage, they fail to consistently capture compositional changes or identify early biomarkers essential for understanding disease progression[10,11]. As a result, clinicians are limited in their ability to identify at-risk joint regions and customize intervention strategies to individual patients[10]. Similarly, pharmacological treatments, including nonsteroidal anti-inflammatory drugs (NSAIDs) and corticosteroids, provide symptomatic relief but do not address the underlying mechanisms driving OA[12,13]. Emerging therapies, including disease-modifying osteoarthritis drugs (DMOADs), show promise but face challenges such as inconsistent efficacy and limited adoption in clinical practice[13,14]. Furthermore, surgical interventions like knee replacement, though effective for late-stage OA, are invasive and impractical for younger or early-stage patients[15,16]. These gaps in current imaging and therapeutic approaches underscore the urgent need for innovative, mechanism-driven solutions to better address OA's complexity and improve patient outcomes.

**Figure 1: Framework for Extracting and Validating Core Components for a Knee Joint Digital Twin.**

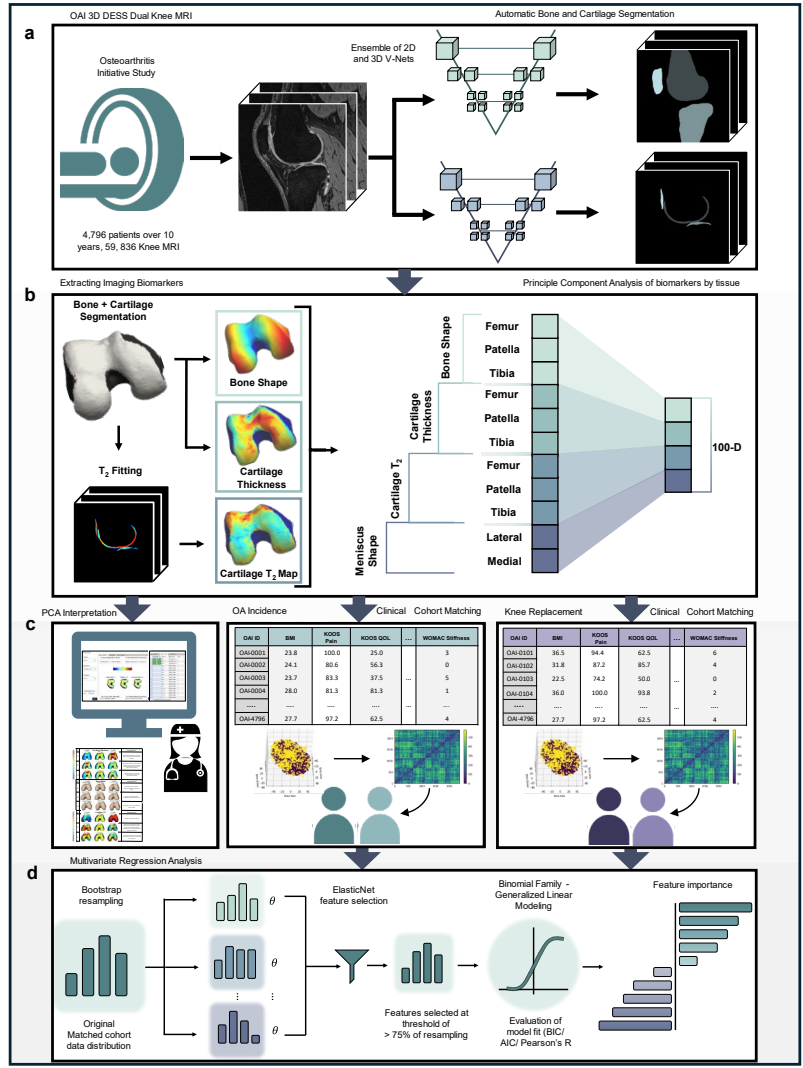

This figure presents a scalable framework for the systematic extraction and analysis of biomarkers from knee joint imaging, laying the groundwork for future digital twin development and studies in osteoarthritis prognosis and intervention.

**a) Imaging Acquisition and Segmentation**: Illustrates the dataset derived from 4,796 unique patients with 3D-DESS MRI scans. Automatic segmentation of bone and cartilage is performed using neural networks, comprising ensembles of 2D and 3D V-Nets previously trained on the OAI dataset. The segmentation captures detailed anatomical structures of the knee joint. The MRI icon in this panel was created by LAFS and obtained from Flaticon.

**b) Biomarker Extraction**: Demonstrates the conversion of segmentation masks into 3D point clouds for bone and meniscus shape analysis. Surface registration is achieved using Iterative Closest Point matching. Cartilage thickness is quantified using Euclidean distance transforms, while $T_2$ relaxation times are calculated from segmented cartilage in MSME-DESS (Multi-Slice Multi-Echo Double Echo Steady State) images using a mono-exponential fitting model. Principal Component Analysis is applied to reduce the dimensionality of these biomarkers, retaining critical geometric variations.

**c) PCA Visualization and Expert Interpretation**: Radiologists use a MATLAB application to visualize and interpret principal component modes, adjusting them within ±3 standard deviations (SDs) from the mean to assess their impact. This is complemented by Clinical Cohort Matching, employing t-SNE embedding and non-parametric testing for cohort analysis in OA Incidence and Knee Replacement studies.

**d) Feature Selection and Regression Analysis**: Elastic net regularization within a Generalized Linear Model framework is applied, iterating over 1,000 bootstrap samples for feature stability. The final logistic regression model evaluates feature importance, calculating coefficients, standard errors, p-values, confidence intervals, and odds ratios to identify significant predictors of OA Incidence and Knee Replacement.

Quantitative MRI (qMRI) has emerged as a valuable innovation[11] with the opportunity to profoundly advance rheumatological care for OA by providing detailed, non-invasive assessments of both structural and compositional changes in joint tissues.[11,17] Unlike conventional MRI, which primarily provides qualitative imaging, qMRI encompasses advanced techniques that yield precise, quantitative measurements of tissue properties.[11] These include morphological features, such as cartilage thickness,[18,19] bone shape,[20–22] and meniscus shape,[23–25] as well as compositional measures like cartilage $T_2$ relaxation time.[26–28] qMRI's ability to capture the structural integrity and biochemical composition of cartilage and surrounding tissue helps us to better understand the underlying mechanisms and progression of OA,[11,17] as well as inform potential interventions.[29]

Recognizing the pressing clinical need and the latent potential of existing technologies, we aim to further enhance the utility of qMRI. This will be achieved by integrating it into a scalable framework that supports precision health through the combination of advanced imaging biomarkers with interpretable and predictive tools. By leveraging qMRI's ability to quantify both structural and compositional changes, we have the opportunity to address current challenges in OA care in the construction of a more personalized approach to disease management. This study represents a foundational step towards the creation of a digital twin for the whole knee joint, designed to predict long-term outcomes such as osteoarthritis and knee replacement. Our intention is to design a comprehensive pipeline (Figure 1) that incorporates advanced imaging techniques, deep learning-based segmentation, and dimensionality reduction in order to build a sustainable and scalable solution[30]. Additionally, this pipeline is designed to mature flexibly with research objectives, with potential for integration of tools like predictive analytics, causal inference, and generative modeling.

Central to this framework is the ability to thoroughly identify and verify risk factors associated with knee health outcomes. This process is supported by the development of an innovative visualization tool; this tool enables the interpretation of our 110-dimensional qMRI feature space in 3D representations, thereby serving as a functional first step toward creating a comprehensible and actionable virtual asset[31] for clinical end-users, facilitating informed decision-making in knee-joint health management.

This study provides a preliminary digital twin framework that leverages advanced qMRI imaging and machine learning techniques to predict the incidence of osteoarthritis and the likelihood of knee replacement. By emphasizing the importance of interpretability and clinical integration, we aim to create an evidenced-based solution[32] that is both scientifically robust and practically valuable. This sets the stage for future longitudinal studies and the broader application of digital twin technology in precision health.

**Results**

*Data Description*

We analyzed imaging and clinical data from the Osteoarthritis Initiative (OAI), a large, multi-center longitudinal study investigating OA progression and outcomes over a decade[33]. The dataset included knee radiographs and dual-knee MRIs from 4,796 participants, collected at baseline and follow-up visits conducted at 12, 24, 36, 48, 72, and 96 months. The imaging protocol incorporated 3D double-echo steady-state (3D-DESS) and 2D multi-slice multi-echo (2D-MSME) MRI scans, complemented by detailed demographic and clinical data such as age, sex, BMI, pain scores, and functional assessments (see Supplementary Data 1).

At baseline, after initial data cleaning steps, the cohort consisted of 4,283 participants. The sex distribution was 58.2% female (n = 2,494) and 41.8% male (n = 1,789). Participants had a median age of 61 years (interquartile range [IQR]: 16 years), encompassing a broad age range from middle-aged to elderly adults. The median BMI was 28.2 kg/m² (IQR: 6.5 kg/m²), indicating that the cohort included individuals from normal weight to obese categories. The dataset incorporated clinical evaluations such as the Knee injury and Osteoarthritis Outcome Score (KOOS)[34], which evaluates pain, physical function, and quality of life related to knee health. At baseline, the KOOS Pain subscale had a median score of 88.9 (IQR: 25), and the

KOOS Quality of Life (QOL) subscale had a median score of 68.8 (IQR: 33.3), indicating varied perceptions of knee-related pain and quality of life among participants. Detailed distributions of the study population's baseline demographic and clinical characteristics are provided in Supplementary Data 2.

*Imaging Biomarker Extraction and Dimensionality Reduction*

MRI data from all subjects underwent advanced neural network-based segmentation to delineate the femoral, tibial, and patellar cartilage, as well as meniscus and bone structures. The segmentation models, validated using a subset of OAI data, achieved Dice coefficients above 0.85 for soft tissues and 0.95 for bones, confirming their accuracy and suitability for imaging biomarker extraction (see Methods Section: *Automatic Knee Joint Tissue Segmentation*).

From these segmentations, we derived a suite of imaging biomarkers that captured structural and compositional variations in the knee joint. These biomarkers included cartilage thickness, cartilage $T_2$ relaxation times, and geometric features of bone and meniscus shapes, which collectively characterize the health and integrity of the joint components. Given the high dimensionality of the extracted data, Principal Component Analysis (PCA) was applied independently to each tissue type and biomarker category. This approach ensured that the most salient variations specific to each tissue and biomarker were retained while maintaining interpretability (Supplementary Data 3).

A total of 110 principal components (PC) modes were computed across tissue and biomarker types, balancing variance capture with practical interpretability. Each tissue-biomarker combination captured 45–80% of the variance within the first 10 PCs, a threshold chosen to support meaningful visualization and clinical interpretation (Supplementary Figure 1). This dimensionality reduction facilitated focused analyses while preserving essential biological variability for downstream investigations.

**Figure 2: Interactive Tool for 3D Visualization and Expert Interpretation of PCA-Derived Biomarkers in the OAI Knee Imaging Feature Space.**

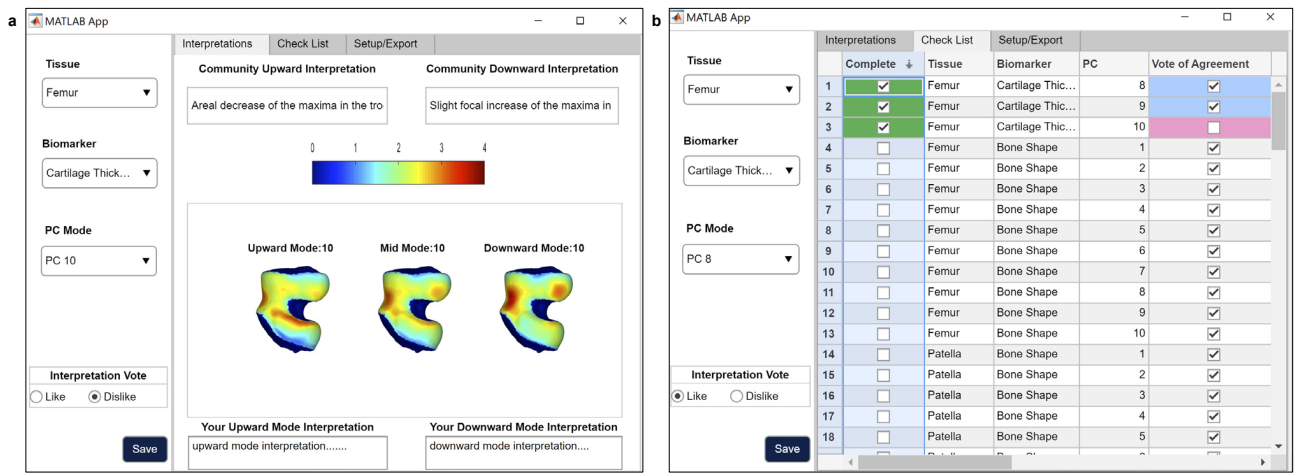

This figure presents an interactive tool designed for radiologists and researchers, enabling the exploration and interpretation of a 3D representation of the OAI Knee Imaging Feature Space within a 110-dimensional Principal Component Analysis framework.

**a) 3D Biomarker Visualization:** This interface allows users to explore the PCA feature space by selecting specific joint tissues and imaging biomarkers. It displays expert radiologist interpretations for notable deviations within ±3 standard deviations from the mean value in the selected PC. The platform supports community engagement by enabling users to contribute their interpretations, interact with community insights, and express confidence in the prevailing interpretations.

**b) Progress Tracking and Community Insight Dashboard:** The 'Check-List' tab within the app serves as a hub for users to track their review of the multidimensional feature space, allowing them to mark the analysis status of various PC modes. It enables filtering by tissue, biomarker, and PC dimension, and presents a consolidated view of community interpretations, capturing both consensus and dissent on insights from the feature space.

*Interactive Visualization and Interpretation of Imaging Biomarkers*

To promote the interpretation of the 110-dimensional PCA feature space, we developed a visualization tool (Figure 2). The visual interface is useful for characterizing structural and compositional tissue changes represented by a PC mode, and considering its clinical applicability. The tool displays the average 3D representation of a joint feature alongside variations at both ends of the distribution curve, simulating the range of values seen within the patient population. Radiologists can examine these feature shifts in order to better understand the variability captured by each mode.

The tool integrates PCA-derived features with their anatomical context, providing a framework for analyzing this complex imaging data. Radiologists used it to evaluate the top 10 PC modes for each biomarker in the baseline data. Tables to summarize mean values and observed variations (mean ± 3 standard deviations) are provided in Supplementary Figures 2-11. These analyses clarify the anatomical and compositional features captured by PCA and qualify the extracted biomarkers as interpretable and clinically meaningful.

**Figure 3: Subject Cohort Selection and Data Processing for OA Incidence and Knee Replacement Analyses.**

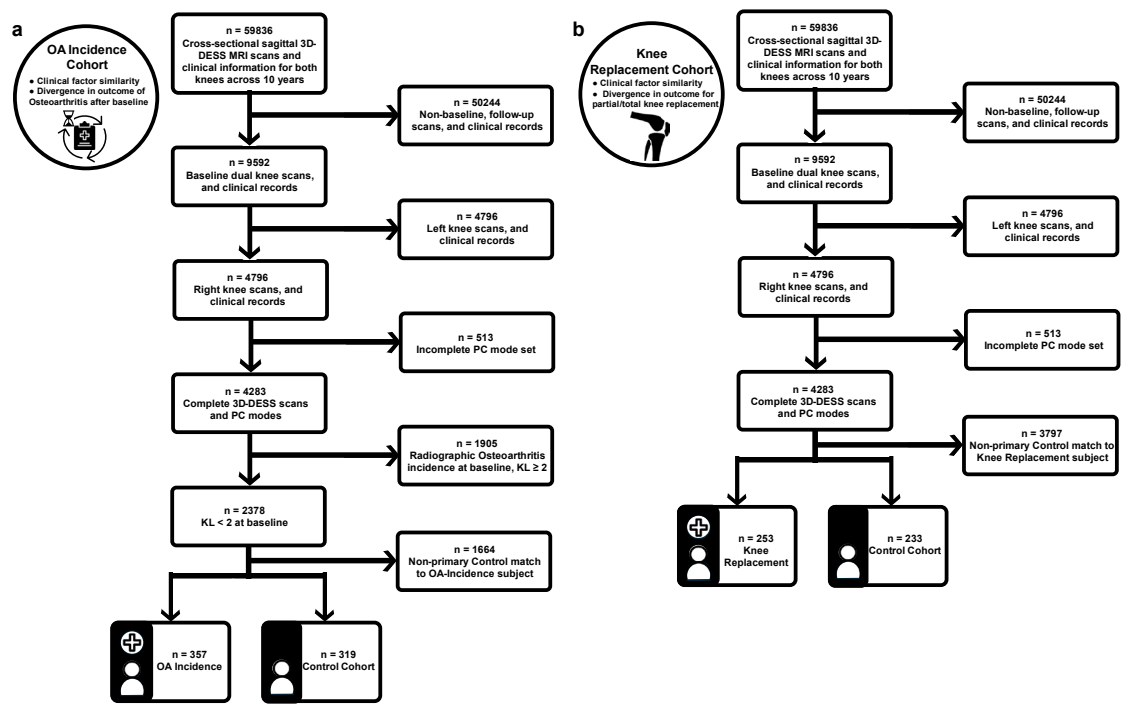

This figure presents flowcharts outlining the steps for cohort selection and data preprocessing, which were used to prepare high-quality datasets for subsequent analyses of OA Incidence and Knee Replacement outcomes in this study.

**a) OA Incidence Cohort Selection and Data Processing:** Illustrates the selection process for the OA Incidence Cohort Analysis, detailing the progression from the initial dataset through various data cleaning stages, including the exclusion of subjects based on the completeness of 3D-DESS MRI scans, clinical records, and Principal Component modes. The final cohort consists of subjects with no initial osteoarthritis (Kellgren-Lawrence Grade [KL] < 2) and their matched controls, all of whom are monitored for OA development over time. The medical chart icon in this panel was created by Kiranshastry and obtained from Flaticon.

**b) Knee Replacement Cohort Selection and Data Processing**: Presents the selection process for the Knee Replacement Cohort Analysis, outlining the data cleaning steps, including the removal of subjects with incomplete data, and the application of stringent inclusion criteria. The final cohort includes subjects who underwent knee replacement surgery and their matched controls who did not.

*Cohort Selection*

To investigate specific knee outcomes, including OA Incidence and Knee Replacement (KR), baseline MRI data were selected according to predefined inclusion and exclusion criteria to ensure data quality and reliability. These criteria involved excluding subjects with incomplete imaging data, missing clinical records, or inadequate scan quality during preprocessing. This approach verified that only subjects with comprehensive and high-quality data were included in the subsequent analyses. We addressed minimal missing data through multiple imputation, followed by sensitivity analyses using Kolmogorov-Smirnov tests to confirm that imputation preserved the original data distributions (Supplementary Data 4 and 5).

Cohort matching was performed to create comparable groups for accurate identification of imaging biomarkers associated with specific knee outcomes, reducing potential confounding effects from demographic and clinical variables. We used t-distributed Stochastic Neighbor Embedding (t-SNE)[35] on standardized demographic and clinical covariates, followed by nearest-neighbor matching in the embedded space (see Methods Section: *Cohort Matching*). This approach was designed to reduce disparities in key variables such as age, BMI, and pain scores between the Outcome and Control groups, thereby enhancing the comparability of the cohorts.

Figure 3 illustrates the selection process, showing the progression from the initial dataset to the finalized cohorts prepared for analysis. In Figure 3a, we detail the OA Incidence cohort selection, comprising subjects with no radiographic OA (Kellgren-Lawrence Grade [KLG] < 2) at baseline who were monitored for OA development over time. Matched controls with similar demographic and clinical profiles were included to ensure a balanced comparison. Figure 3b depicts the KR cohort, comprising participants who underwent knee replacement surgery and their matched controls.

Post-matching, the OA Incidence cohort consisted of 357 cases and 319 matched controls. The median age was 60 years in both groups, and the median BMI was approximately 29 kg/m². KOOS Pain subscale scores were similar between groups, with medians of 90.6 (IQR: 22.2) for the OA Incidence group and 91.7 (IQR: 19.4) for the Control group, indicating comparable levels of knee pain. The KOOS Quality of Life (QOL) scores were slightly lower in the OA Incidence group (median 68.8, IQR: 37.5) compared to the Control group (median 75.0, IQR: 31.2), suggesting a minor difference in knee-related quality of life (Supplementary Data 6).

In the KR cohort, the KR group included 253 participants to the Control group's 233 participants. Both the KR and Control groups had a median age of 64 years and a median BMI around 29 kg/m². KOOS Pain subscale scores were slightly lower in the KR group (median 72.2, IQR: 27.8) compared to the Control group (median 75.0, IQR: 25.4), suggesting marginally worse knee pain among those who underwent knee replacement. The KOOS QOL scores were similar between groups, with medians of 56.3 (IQR: 25.0) in the KR group and 56.3 (IQR: 31.3) in the Control group, reflecting greater knee dysfunction compared to the OA Incidence cohort (Supplementary Data 7).

These demographic and clinical characteristics establish a baseline profile for the matched cohorts. This alignment supports comparative analyses and strengthens the identification of imaging biomarkers associated with knee health outcomes.

*Cohort Matching Analytical Integrity*

To ensure that our matched cohorts were appropriately balanced and that any observed differences in imaging biomarkers were not due to confounding variables, we evaluated the analytical integrity of our cohort matching. The findings revealed that the matching technique reduced disparities across demographic and clinical variables, thereby improving the validity of subsequent analyses.

The comparability of the matched cohorts was evaluated using a series of statistical measures; we employed Standardized Mean Difference (SMD) and Cohen's d to quantify differences in continuous variables and Cramer's V for categorical variables between the Outcome and Control groups. Additionally, we calculated the Point Biserial Correlation Coefficient to assess the strength of association between each covariate and group classification. To evaluate the statistical significance of any remaining disparities, we applied non-parametric hypothesis tests, including the Wilcoxon signed-rank test for paired continuous variables and the Chi-squared test for categorical data.

For the OA Incidence analysis, the matching process substantially improved the balance between the OA Incidence and Control groups. Figure 4a illustrates the significant reductions in SMD values for numerical covariates following matching, with values shifting from higher levels (teal) to lower levels (purple), reflecting improved covariate balance. Key variables such as age, BMI, and KOOS scores showed marked reductions in disparities after matching. Specifically, Cohen's d values decreased significantly (see Supplementary Data 8); for example, the Cohen's d for BMI reduced from -0.528 before matching to -0.053 after matching, demonstrating a substantial improvement in balance. Additionally, reduced associations with group classification were observed, as the Point Biserial Correlation coefficients became nonsignificant for the majority of variables.

Non-parametric tests further confirmed the success of the matching process. Results from the Wilcoxon signed-rank tests indicated that, for the majority of continuous variables, no significant differences remained between the Outcome and Control groups after matching. Minor significant differences remained for weight and BMI, with p-values of 0.025 and 0.036, respectively (see Supplementary Data 9). Despite these residual disparities, the effect sizes were greatly diminished compared to pre-matching, and the absolute differences were minimal. Figure 4c depicts the distribution of categorical covariates between the Control and OA Incidence groups post-matching, demonstrating improved balance for variables such as sex and race.

For the Knee Replacement analysis, the matching process effectively minimized disparities across most variables, as reflected by SMD and Cohen's d values approaching zero. As shown in Figure 4b, SMD assessments for numerical covariates illustrate the effectiveness of the matching process, with effect sizes decreasing from pre-match (teal) to post-match (purple), demonstrating improved equivalence between cohorts (see Supplementary Data 10). To provide an example, the Cohen's d for KOOS Pain presented a substantial reduction in the disparity, with a decrease in magnitude from -0.649 to 0.015 post-matching. Similarly, the Point Biserial Correlation indicated weaker associations, with smaller coefficients and non-significant p-values.

Despite these improvements, the Chi-squared tests indicated that some categorical variables, such as race and Hispanic ethnicity, still exhibited significant differences post-matching. Specifically, the Chi-squared test for race yielded a statistic of 9.81 with a p-value of 0.020 (see Supplementary Data 11), suggesting that perfect balance was not achieved for this variable. While Cramer's V values indicate lingering disparities for some categorical variables, Figure 4d shows that the matching process effectively aligned the group characteristics, with only slight remaining differences.

In this way, this evaluation demonstrates the effectiveness of our cohort matching. By improving balance across demographic and clinical variables, the process reduces confounding effects and strengthens associations between imaging biomarkers and knee health outcomes. Residual differences are minimal and unlikely to impact results.

**Figure 4: Evaluating Matching Quality and Identifying Significant Biomarkers in OA Incidence and Knee Replacement Cohorts.**

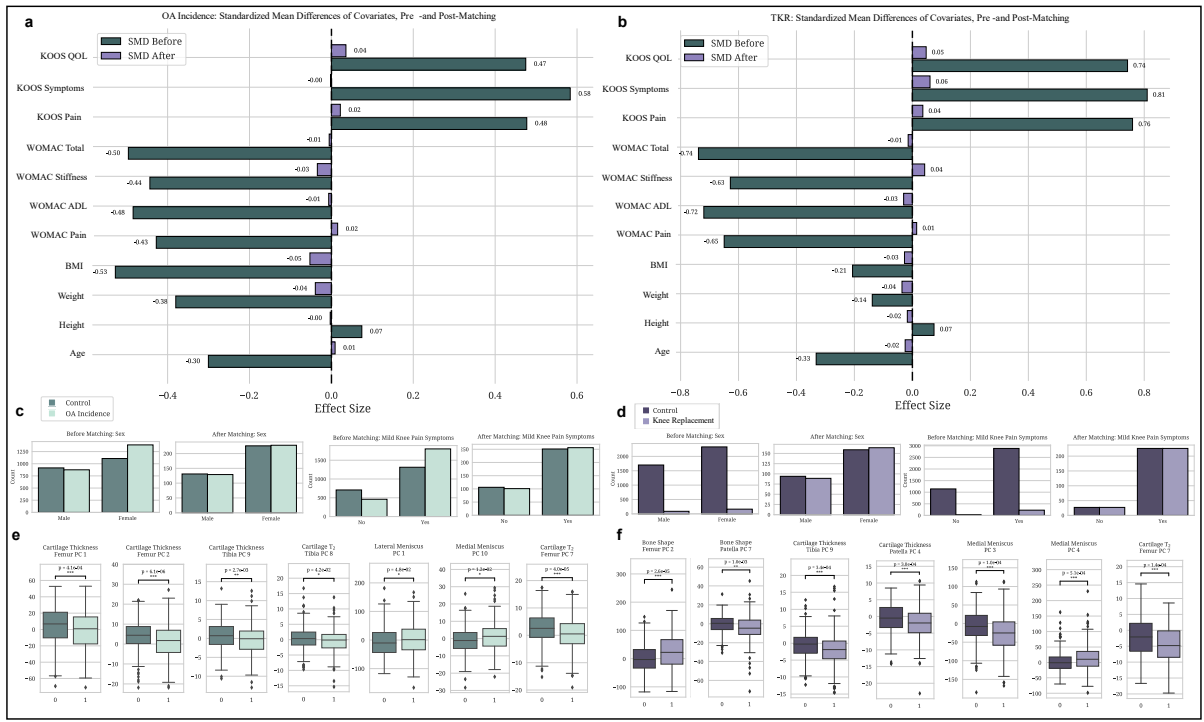

This figure evaluates the quality of cohort matching and accentuates significant differences in imaging biomarkers between the Control and Outcome groups for OA Incidence and Knee Replacement analyses.

**a) OA Incidence Matching Quality**: Standardized Mean Difference (SMD) analysis compares numerical covariates before (teal) and after (purple) matching, visualizing improved balance (see Supplementary Data Table 8).

**b) Knee Replacement Matching Quality**: The SMD assessment for numerical covariates illustrates the effectiveness of the matching process. Effect sizes are depicted in teal for pre-match and purple for post-match conditions, demonstrating improved equivalence between cohorts (see Supplementary Data Table 10).

**c) OA Incidence Covariate Balance**: Bar plots depict the distribution of categorical covariates between Control (dark teal) and OA incidence (light teal) groups. This analysis highlights the improvement in balance for categorical variables, such as Sex (see Supplementary Data Table 9).

**d) Knee Replacement Covariate Balance**: Bar plots illustrate the distributions of categorical covariates for Control (dark purple) versus Knee Replacement (light purple) groups. These plots show the success of the matching process in equalizing group characteristics (see Supplementary Data Table 11).

**e) Significant Features in OA Incidence Cohort Study**: Box plots display the distribution of significant PC mode features between Control and OA incidence groups. Box plots display the median, interquartile range, and whiskers extending to 1.5 times the IQR. Statistical significance was determined using Paired Wilcoxon Rank Sum Tests with Benjamini-Hochberg correction for multiple comparisons (p-values adjusted using the Hochberg method). Significance levels are denoted as follows: single asterisk for $p < 0.05$, double asterisks for $p < 0.01$, and triple asterisks for $p < 0.001$, aligning with results in Supplementary Data Table 14.

**f) Significant Features in Knee Replacement Cohort Study**: Box plots, as described in panel e, compare significant PC mode features between Control and Knee Replacement groups. Statistical significance was determined using Paired Wilcoxon Rank Sum Tests with Hochberg correction. Annotated p-values and significance thresholds are provided in Supplementary Data Table 15. Remaining significant PC mode plots not shown are in Supplementary Figures 12 and 13.

*Cross-Sectional Analyses of Imaging Biomarkers for OA Incidence and Knee Replacement*

Building upon the matched cohorts, we conducted cross-sectional analyses to identify significant differences in imaging biomarkers between the Outcome and Control groups. Prior to these analyses, normality and variance of PC modes were evaluated using Shapiro-Wilk and Anderson-Darling tests to confirm statistical validity (see Supplementary Data 12 for OA Incidence and Supplementary Data 13 for Knee Replacement). We examined the 110 PC modes relative to each cohort and sought to elucidate the associations between the imaging biomarkers and the OA incidence and KR outcomes. By determining the joint features indicative of downstream knee outcomes, we initiate a solid basis for a comprehensive digital twin of the knee joint; these imaging biomarkers are the constituents of the model from which precise tracking and simulation of disease progression may be derived.

For the OA Incidence cohort, analyses using the Wilcoxon rank-sum test with Benjamini-Hochberg correction ($\alpha = 0.05$) revealed significant differences in imaging biomarkers between the OA Incidence and Control groups. The cohort included 357 cases and 319 matched controls, and the analyses focused on structural and compositional features of the knee joint, including cartilage thickness, cartilage $T_2$ relaxation times, and meniscus shape.

Cartilage Thickness Femur PC1 and PC2 showed significant decreases in the OA Incidence group compared to controls, with adjusted p-values of $4.12 \times 10^{-4}$ and $6.11 \times 10^{-6}$, respectively. This indicates thinner cartilage in critical regions of the femur among individuals who developed OA; this suggests that early cartilage degeneration in the femur may be a key indicator of OA onset. Similarly, significant decreases were observed in Cartilage Thickness Tibia PC9 (adjusted p-value = 0.00273), which reflects cartilage thinning in the tibial region.

Significant alterations were also found in cartilage composition. Cartilage $T_2$ Femur PC7 (adjusted p-value = $3.96 \times 10^{-5}$) and Cartilage $T_2$ Tibia PC8 (adjusted p-value = 0.0416) exhibited significant decreases in the OA Incidence group. Changes in cartilage $T_2$ relaxation times suggest variations in the biochemical properties of cartilage, such as collagen content and water distribution, which are essential for cartilage health.

Furthermore, significant differences were identified in meniscus shape PC modes. Medial Meniscus PC10 (adjusted p-value = 0.01195) and Lateral Meniscus PC1 (adjusted p-value = 0.04837) showed increases in the OA Incidence group compared to controls. Alterations in meniscal morphology may contribute to the mechanical environment that promotes OA progression. These findings reveal specific imaging biomarkers associated with the early stages of OA, and stress the importance of both cartilage and meniscal changes in OA development.

Similar analyses revealed significant differences in PC modes related to bone shape, cartilage thickness, cartilage $T_2$ relaxation times, and meniscus shape for the Knee Replacement cohort. The cohort included 253 subjects who ultimately underwent a partial or total knee replacement to remedy their failing joint, and 233 matched control subjects who did not require surgical intervention.

Bone Shape Femur PC2 exhibited a significant increase in the KR group compared to controls (adjusted p-value = $2.55 \times 10^{-5}$), which indicates alterations in femoral bone morphology associated with advanced joint degeneration. This demonstrates the importance of bone structural changes in the progression to joint dysfunction that necessitates surgical intervention. In contrast, Bone Shape Patella PC7 showed a significant decrease in the KR group (adjusted p-value = 0.00100), suggesting changes in patellar bone structure.

Significant decreases were observed in Cartilage Thickness Femur PC2 (adjusted p-value = $2.01 \times 10^{-5}$) and Tibia PC2 (adjusted p-value = $2.01 \times 10^{-5}$); this reflects substantial cartilage loss in key load-bearing regions of the knee joint. These structural changes contribute to joint dysfunction and pain experienced by individuals requiring knee replacement. Additionally, Cartilage Thickness Femur PC3 showed a significant increase (adjusted p-value = $1.33 \times 10^{-4}$), which indicates regional variations in cartilage thickness alterations.

In terms of cartilage composition, Cartilage $T_2$ Femur PC5 demonstrated a significant increase in the KR group (adjusted p-value = $1.18 \times 10^{-5}$), while Cartilage $T_2$ Femur PC7 exhibited a significant decrease (adjusted p-value = $1.38 \times 10^{-4}$). Similarly, Cartilage $T_2$ Tibia PC1 showed a significant increase (adjusted p-value = $1.04 \times 10^{-4}$). These findings reveal degradation of cartilage biochemical properties associated with more advanced disease stages.

Significant differences were also found in meniscus shape PC modes, particularly in the medial meniscus. Medial Meniscus PC3 (adjusted p-value = $1.04 \times 10^{-4}$) and Medial Meniscus PC8 (adjusted p-value = $2.97 \times 10^{-4}$) showed significant decreases in the KR group. This demonstrates the role of meniscal degeneration in the progression of knee joint instability to the point of requiring replacement surgery. Conversely, Medial Meniscus PC4 exhibited a significant increase (adjusted p-value = $5.06 \times 10^{-4}$), which suggests complex alterations in meniscal morphology.

*Integrated Findings from Cross-Sectional Analysis*

The analyses of the OA Incidence and KR cohorts identified distinct structural and compositional changes in the knee joint associated with disease progression and surgical outcomes. In the OA Incidence cohort, early changes were observed, including significant decreases in cartilage thickness, alterations in cartilage composition, and morphological differences in meniscus shape. These biomarkers were consistently associated with the early stages of knee degeneration.

In parallel, analyses of the Knee Replacement cohort recognized meaningful biomarkers that reflected advanced joint degeneration. Substantial cartilage loss in key load-bearing regions, alterations in bone morphology, and meniscal degeneration were notable features distinguishing individuals requiring knee replacement surgery from matched controls. Bone shape and meniscus morphology were prominent contributors to group differences; this observable divergence is specific to structural changes associated with advanced disease stages.

Collectively, these biomarkers may provide key insights into modeling knee joint health and tracking disease progression dynamically. Supporting statistical results are detailed in Supplementary Data 14 and 15, with visual summaries in Figure 4e-f, Supplementary Figures 12 and 13.

*Multivariate Regression Analysis for OA Incidence and Knee Replacement Cohorts*

In order to extend the cross-sectional findings, we utilized multivariate regression to determine new imaging biomarkers of value and validate the importance of those previously identified. This approach functionally assessed the independent contributions of biomarkers to knee health outcomes while accounting for potential confounders. Consequently, these results allow us to better understand structural and compositional factors of the joint; they support the creation of a digital twin framework for dynamic disease modeling and personalized monitoring.

We applied multivariate regression for the OA Incidence cohort using elastic net regularization within a generalized linear model (GLM). Elastic net regularization was chosen to manage multicollinearity among the numerous PC modes while performing feature selection, and stability selection ensured that the identified predictors were robust across different sample subsets (see Supplementary Data 16 for raw data). The analysis incorporated 110 PCs in addition to key demographic and clinical covariates such as age, BMI, and pain scores. Significant predictors determined based on their weighted importance scores and inclusion frequency across bootstrap iterations (Figure 5a, Supplementary Figures 14a-d, 15, and Supplementary Data 17).

**Figure 5: Stability Selection and Distribution of Imaging Biomarkers from Multivariate Analysis of OA Incidence and Knee Replacement Outcomes.**

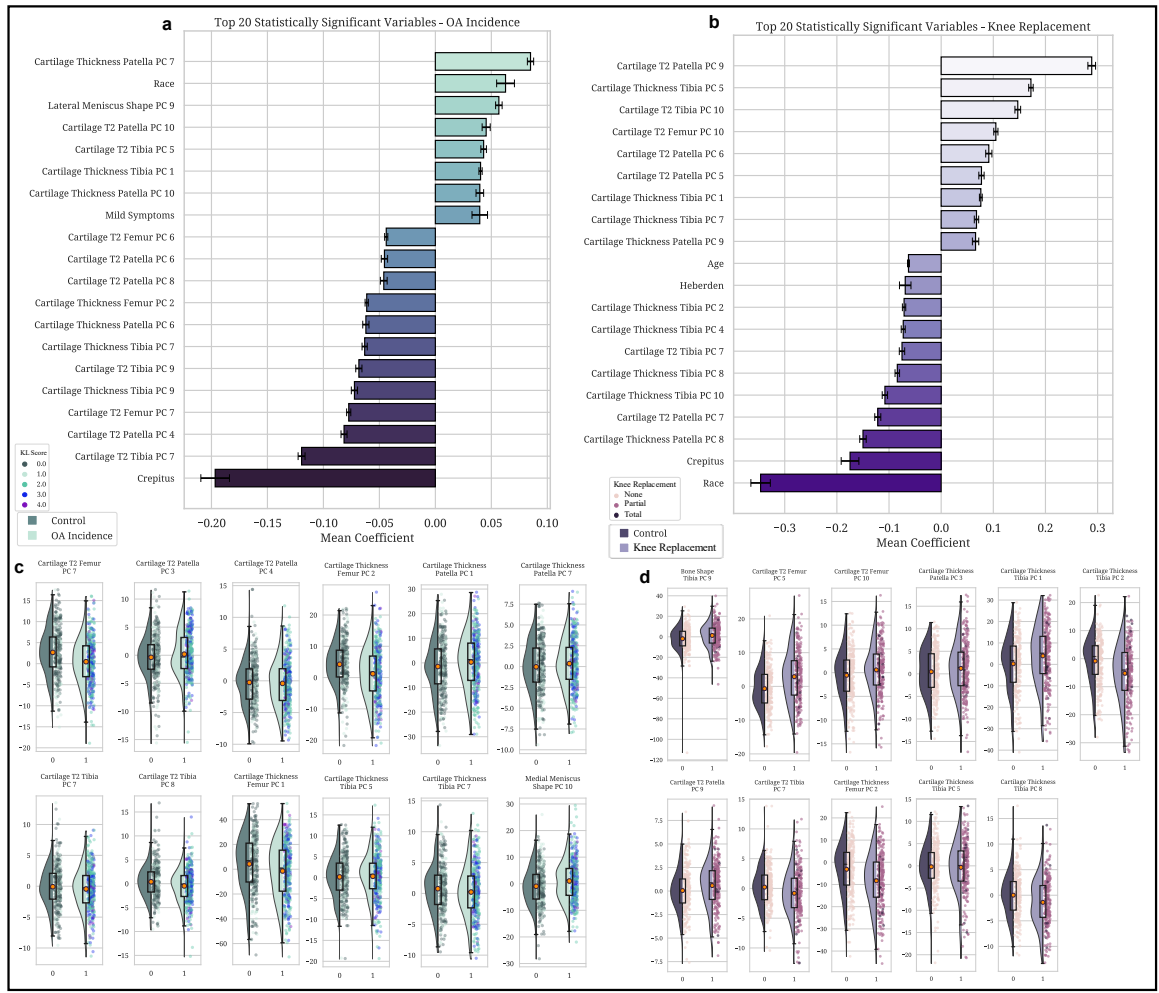

This figure presents imaging biomarkers identified through stability selection and multivariate analysis. These features have the potential to inform knee health dynamics and guide further exploration into biomarkers relevant to digital twin development.

**a) OA Incidence Stability Selection – Overall Top 20 Features**: This horizontal bar chart displays the top 20 significant variables associated with OA incidence, ranked by mean coefficients derived from bootstrap elastic net analysis over 1,000 iterations. Positive effects extend to the right, while negative effects extend to the left, with 95% confidence intervals indicated by horizontal lines. Detailed results are available in Supplementary Data Table 17 and visualized by imaging biomarker category in Supplementary Figure 14a-d.

**b) Knee Replacement Stability Selection – Overall Top 20 Features**: Presents a horizontal bar chart of the top 20 significant variables for knee replacement outcomes, also ranked by mean coefficients from bootstrap elastic net analysis with 1,000 iterations. Positive effects extend to the right, negative effects to the left, with 95% confidence intervals indicated. Additional details are in Supplementary Data Table 20 and categorized views are shown in Supplementary Figure 14e-h.

**c) Distribution Plots of PC modes for OA Incidence**: Features violin plots of significant PC modes from the final GLM logistic regression, comparing OA incidence (light teal) and Control groups (dark teal). The violin plots are superimposed with dot plots representing Kellgren-Lawrence (KL) grades, a radiographic measure of OA severity (KL grade ≥ 2 indicates radiographic OA). Dot plots reflect ultimate knee outcomes over time in the OAI study. Statistical details are provided in Supplementary Data Table 18 and further visualized in Supplementary Figure 15.

**d) Distribution Plots of PC modes for Surgical Intervention:** Violin plots display significant PC modes from the final GLM logistic regression, comparing Control (dark purple) and Knee Replacement (light purple) groups, further stratified by non-replacement, partial replacement, and total knee replacement. These dot plots reflect ultimate knee outcomes over time in the OAI study. Additional statistical details can be found in Supplementary Data Table 21 and further visualized in Supplementary Figure 16.

Several PC modes related to cartilage thickness and cartilage T$_2$ relaxation times emerged as significant predictors of OA incidence (Figure 5c, Supplementary Data 18). Specifically, higher scores in Cartilage T$_2$ Femur PC7 (odds ratio [OR] = 0.949; 95% confidence interval [CI]: 0.918–0.982; $p = 2.35 \times 10^{-3}$) and Cartilage Thickness Femur PC1 (OR = 0.973; 95% CI: 0.959–0.987; $p = 1.82 \times 10^{-4}$) were associated with decreased odds of OA incidence. This suggests that preserved structural and compositional characteristics in femoral cartilage may protect against early joint health decline. Similarly, Cartilage Thickness Femur PC2 demonstrated a protective role (OR = 0.960; 95% CI: 0.935–0.985; $p = 2.05 \times 10^{-3}$), further reinforcing its relevance as a marker of resilience in joint health and a candidate for dynamic tracking within a digital twin framework.

Conversely, higher scores in Cartilage T$_2$ Patella PC3 (OR = 1.049; 95% CI: 1.001–1.099; $p = 4.37 \times 10^{-2}$) and Cartilage Thickness Patella PC7 (OR = 1.061; 95% CI: 1.004–1.121; $p = 3.61 \times 10^{-2}$) were associated with increased odds of OA incidence, indicating that specific patellar cartilage changes may signify an elevated risk of degeneration. Meanwhile, higher scores in Cartilage T$_2$ Patella PC4 (OR = 0.950; 95% CI: 0.902–0.999; $p = 4.68 \times 10^{-2}$) were associated with decreased odds of OA incidence that preserved biochemical properties of the patellar cartilage could mitigate risk.

In the tibial cartilage, Cartilage T$_2$ Tibia PC7 (OR = 0.935; 95% CI: 0.888–0.985; $p = 1.10 \times 10^{-2}$) and Cartilage T$_2$ Tibia PC8 (OR = 0.944; 95% CI: 0.898–0.992; $p = 2.31 \times 10^{-2}$) showed protective effects against OA incidence; this emphasizes the importance of maintaining tibial cartilage composition in preventing early disease onset. These findings further support the utility of these biomarkers for dynamic simulation and monitoring within the digital twin framework. Additionally, Cartilage Thickness Tibia PC7 demonstrated a protective role (OR = 0.943; 95% CI: 0.903–0.984; $p = 7.44 \times 10^{-3}$), while Cartilage Thickness Tibia PC5 was associated with increased OA risk (OR = 1.039; 95% CI: 1.002–1.079; $p = 4.10 \times 10^{-2}$). These findings emphasize structural variations that may predispose individuals to degeneration.

Additionally, Medial Meniscus PC10 was modestly associated with increased OA incidence odds (OR = 1.026; 95% CI: 1.002–1.051; $p = 3.28 \times 10^{-2}$). This indicates that morphological changes in the meniscus may influence the mechanical environment of the knee joint and contribute to the early stages of OA development.

In the Knee Replacement cohort, multivariate regression revealed significant predictors of advanced joint degeneration requiring surgical intervention. The analysis utilized the same 110 PCs to discern factors contributing to advanced joint instability (Figure 5b, Supplementary Figures 14e-h, 16, and Supplementary Data 19 and 20). Significant predictors included PC modes related to bone shape, cartilage thickness, and cartilage T$_2$ relaxation times (Figure 5d, Supplementary Data 21).
Bone Shape Tibia PC9 was associated with increased likelihood of knee replacement (OR = 1.038; 95% CI: 1.016–1.059; $p = 4.82 \times 10^{-4}$). This highlights tibial bone morphological changes that contribute to advanced joint instability and may serve as an indicator for surgical intervention. Conversely, Cartilage T$_2$ Tibia PC7 displayed a protective effect (OR = 0.890; 95% CI: 0.833–0.952; $p = 6.18 \times 10^{-4}$), suggesting that preserved cartilage composition in specific tibial regions mitigates the progression toward severe joint degeneration.

In tibial cartilage thickness, Cartilage Thickness Tibia PC1 (OR = 1.034; 95% CI: 1.014–1.054; $p = 7.52 \times 10^{-4}$) and PC5 (OR = 1.060; 95% CI: 1.012–1.111; $p = 1.41 \times 10^{-2}$) were associated with increased risk, which potentially reflects structural changes in load-bearing areas that exacerbate joint degradation. In contrast, PC2 (OR = 0.939; 95% CI: 0.911–0.968; $p = 4.58 \times 10^{-5}$) and PC8 (OR = 0.932; 95% CI: 0.884–0.983; $p = 9.40 \times 10^{-3}$) showed protective effects, indicating that specific features of tibial cartilage integrity might reduce susceptibility to severe outcomes.

In the femoral cartilage, Cartilage T$_2$ Femur PC5 (OR = 1.064; 95% CI: 1.027–1.102; $p = 6.11 \times 10^{-4}$) and PC10 (OR = 1.073; 95% CI: 1.026–1.122; $p = 2.04 \times 10^{-3}$) were significant risk factors; this suggests that compositional changes may mark advanced stages of degeneration. Contrastingly, Cartilage Thickness Femur PC2 (OR = 0.948; 95% CI: 0.927–0.970; $p = 3.88 \times 10^{-6}$) demonstrated a protective role, reinforcing its importance as a key metric for structural integrity and dynamic tracking within the digital twin framework.

Furthermore, Cartilage Thickness Patella PC3 (OR = 1.062; 95% CI: 1.015–1.110; $p = 8.69 \times 10^{-3}$) and Cartilage $T_2$ Patella PC9 (OR = 1.145; 95% CI: 1.041–1.259; $p = 5.31 \times 10^{-3}$) were associated with increased risk of knee replacement. This emphasizes the contribution of patellar cartilage alterations to joint dysfunction in advanced stages of degeneration.

*Integrated Findings from Multivariate Regression*

The multivariate regression analyses identified imaging biomarkers that independently correlate with OA incidence and knee replacement outcomes; said biomarkers reveal distinct structural and compositional characteristics of the knee joint. In the OA Incidence cohort, early changes were characterized by protective features such as preserved femoral and tibial cartilage thickness and composition. In contrast, the Knee Replacement cohort showed advanced degeneration markers; alterations appeared in bone morphology and cartilage in key load-bearing regions.

Together, these results provide a comprehensive basis for understanding the progression of joint degeneration across disease stages. By characterizing both early resilience and advanced deterioration, these findings establish a framework for dynamic monitoring and modeling of joint integrity; as such, the recognized features have potential application in digital twin systems to track and forecast knee health dynamics. Supporting statistical results are detailed in Supplementary Data 18 and 21, with visual summaries in Figure 5a-d.

*Core Biomarker Indicators for Digital Twin Development*

Integrating the findings from the cross-sectional cohort analyses and multivariate regression models, we identified imaging biomarkers consistently linked to knee health outcomes across multiple analytical approaches. These biomarkers demonstrated potential reliability as markers of osteoarthritis incidence and predictors of the need for knee replacement surgery. Their consistent associations suggest their potential for integration into a digital twin framework for modeling knee joint health.

In the OA Incidence cohort, biomarkers such as Cartilage Thickness Femur PC1 and PC2, Cartilage $T_2$ Femur PC7, and Tibia PC8 emerged as protective factors, while Medial Meniscus PC10 was associated with increased OA risk. Similarly, in the KR cohort, Cartilage Thickness Femur PC2 demonstrated a protective role, with other biomarkers, including Cartilage $T_2$ Femur PC5 and Bone Shape Tibia PC9, linked to increased surgical risk. The identification of Cartilage Thickness Femur PC2 as a protective factor across both cohorts and analyses suggests it could serve as a potential key metric for tracking femoral cartilage integrity within a digital twin framework.

Figure 6 presents the main biomarkers discovered through our analyses and their association with OA progression and knee replacement outcomes; these biomarkers show potential utility in the early detection of OA, monitoring advanced joint degeneration, and understanding protective effects. In this way, these markers may aptly serve as a basis for risk assessment and the development of a digital twin model for personalized knee health monitoring.

**Figure 6: Core Biomarkers Identified Across Analytical Approaches for OA Incidence and Knee Replacement Outcomes in Digital Twin Development.**

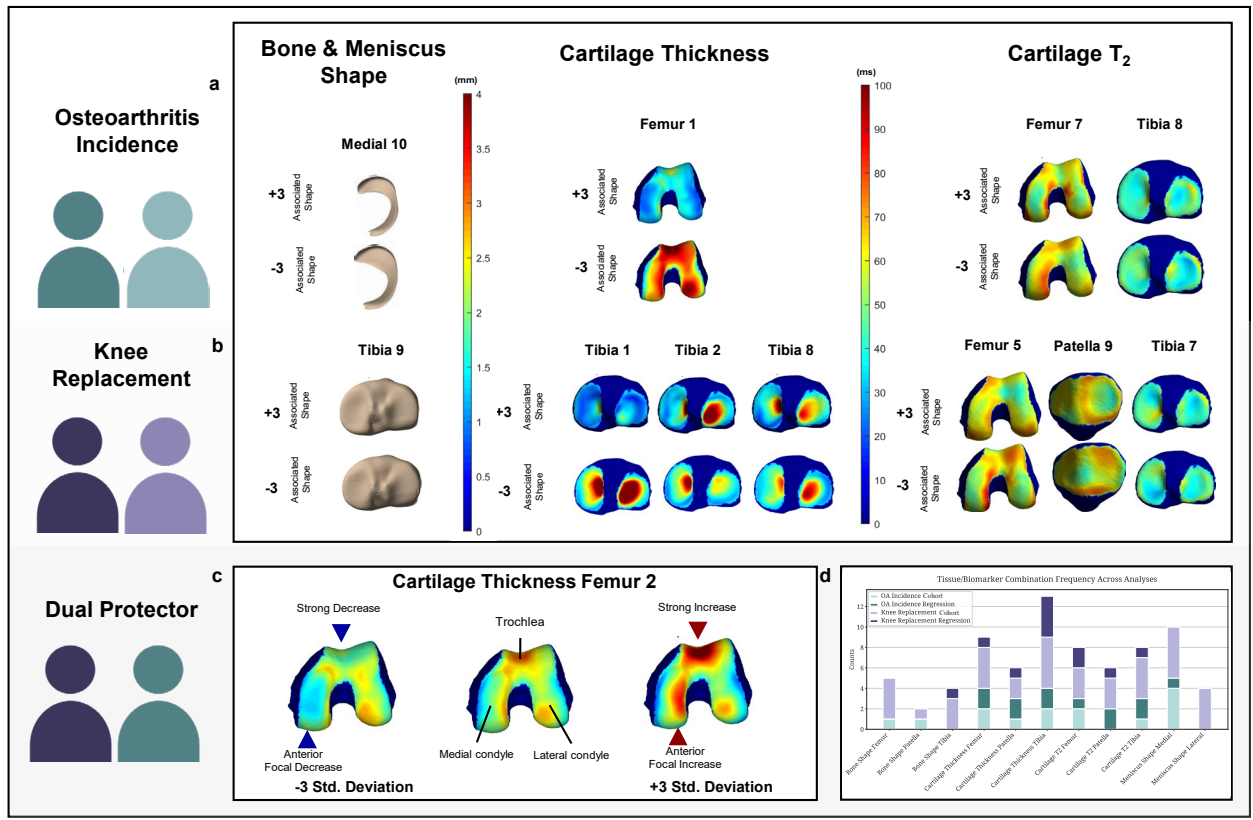

This figure integrates findings from cohort matching and multivariate regression analyses, identifying imaging biomarkers consistently associated with OA incidence and knee replacement outcomes, which serve as foundational components for building a knee joint digital twin system.

**a) OA Incidence Biomarkers**: Emphasizes biomarkers significantly associated with the incidence of osteoarthritis, identified through Clinical Cohort Matching and Multivariate Regression Analyses. Emphasizes specific anatomical and molecular changes linked to the early detection and progression of OA.

**b) Knee Replacement Biomarkers**: Presents biomarkers that distinguish between Control subjects and those who underwent knee replacement surgery. These biomarkers were identified as the most impactful for predicting surgical outcomes through cohort matching and regression.

**c) Dual Protective Biomarker**: Features a biomarker that consistently demonstrates protective effects against both the onset of OA and the need for knee replacement. Based on cohort comparisons and multivariate models, the findings suggest potential therapeutic intervention targets.

**d) Biomarker Selection Overview**: Synthesizes the tissue-biomarker combinations identified as significant across analyses for both OA incidence and knee replacement. The repeated recognition of these biomarkers across analyses accentuates their influence in knee health outcomes and posits their utility for further investigation.

## Discussion

This study represents a foundational step toward building a scalable framework to investigate osteoarthritis incidence and knee replacement risk through data-centric methods that determine and authenticate cardinal imaging biomarkers. Through a robust pipeline that includes dimensionality reduction, statistical cohort matching, and multivariate regression, we extracted clinically relevant imaging biomarkers from a high-dimensional MRI dataset. These biomarkers provide valuable information about knee joint health and lay the groundwork for forming predictive models that guide clinical decision-making, as well as for advancing the concept of a digital twin for the knee joint.

Central to our approach was the establishment of an embedded representation space (a lower-dimensional abstraction derived from the original high-dimensional qMRI feature space); this embedded space encapsulates high-fidelity information for each subject, enabling robust comparisons between Knee Outcome and Control groups and enhancing the reliability of our biomarker findings. We matched subjects based on demographics and clinical factors. After confirming the quality of these matches, we identified key imaging features, such as changes in cartilage thickness, T2 relaxation times, and bone and meniscus shapes, that showed significant associations with OA and KR outcomes. These insights enhance our understanding of OA incidence and have the potential to improve diagnostic accuracy and treatment planning.

In our Cohort Analyses, we identified significant imaging features across our OA Incidence and KR groups, with seven and thirty-eight features, respectively, showing statistical significance. Multivariate regression further substantiated five principal component mode features impacting OA Incidence and three affecting Knee Replacement outcomes. These analyses converged to reveal salient biomarkers, consistent across both cohorts, as detailed in Figure 6.

Namely, localized cartilage thickening in regions such as the trochlea and anterior medial condyle (Femur PC2), demonstrated a protective effect against OA, while generalized thinning correlated with increased OA incidence—findings that echo previous research on the significance of non-stable thickness trajectories in OA.[36] Additionally, alterations in $T_2$ relaxation times were identified as indicative of resistance to OA; decreased $T_2$ in the medial weight-bearing regions of the femur aligned with the work of Baum et al.,[37] while increased $T_2$ at the periphery of the tibia's lateral facet supported previous observations of "regionally heterogenous"[38] cartilage degeneration prior to lesion onset. Furthermore, specific changes in meniscal morphology, such as smaller posterior horns in the medial meniscus, accentuating the role of meniscal integrity in knee health, corroborating earlier studies by Gao et al.[25] and Kawahara et al.[39] that emphasize the vulnerability of the medial posterior horn to OA progression.

Similarly, for knee replacement prognosis, the preservation of cartilage thickness at Femur PC2 and particular tibial regions (Tibia PC2 and PC8) was associated with a lower likelihood of knee replacement; such denotes the importance of localized cartilage resilience in joint integrity[40]. Additionally, bone shape changes, particularly in the tibia (Bone Shape Tibia PC9), were strongly associated with increased surgical intervention risk. This aligns with research emphasizing the uniformity of tibial alterations and localized changes in the femur,[41] suggesting a complex relationship between bone morphology and the course of OA. These findings promote the evidence that 3D bone shape metrics are indeed reliable indicators of OA progression and surgical interventional need.[22,42]

An integral aspect of this study is the construction of a dynamic system capable of handling large imaging datasets for biomarker extraction and validation. This infrastructure supports our immediate objectives and concurrently positions us for extended longitudinal inquiry. In doing so, we will introduce multimodal data and implement advanced causal inference techniques. To this end of exploring the long-term progression of knee outcomes in the OAI dataset, we will project high-dimensional qMRI features from subsequent timepoints onto the baseline PCA space established in our initial analysis. This approach will produce consistent feature embeddings at multiple intervals, allowing for dynamic three-dimensional modeling and visualization.

Equally important is the development of a tool for such visualization to complement our pipeline, which enables the interpretation of the 110-dimensional PC feature set and biomarkers of interest. This tool translates complex imaging data into three-dimensional models that are intuitive and clinically actionable. Beyond furnishing accessible visual representations, it encourages expert engagement by allowing users to review principal component deviations, contribute interpretations, and draw on collective insights. Its progress-tracking capabilities promote organized, group-driven examination of various PC modes; in doing so, it ensures that subtle imaging characteristics receive proper attention. These design choices prioritize explainability and systematic analysis, aiming to deliver predictive analytics that remain accessible to clinical end-users. Notwithstanding, we acknowledge certain limitations of this approach. The variance captured among principal components varied, with bone and meniscus shape modes exceeding 80% variance capture, while cartilage thickness and T2 modes captured between 45-65%; indeed, this illustrates the inherent difficulty in combining interpretability with predictive strength. Nonetheless, this interactive and community-driven approach demonstrates the value of bringing such design principles into biomedical tools; these principles raise the clinical applicability of findings and encourage new directions in research and healthcare solutions.

As we move forward, our immediate focus will be on expanding the pipeline to incorporate additional multimodal data sources, such as biomechanical factors[43] and clinical history, to augment the predictive value of our digital twin framework. This approach is designed with clinical integration in mind; therefore, it has the potential to facilitate the automatic embedding, comparison, visualization, and interpretation of patient data to monitor knee health progression. To further this objective, we will apply our framework to larger and more diverse cohorts, in order to strengthen its generalizability and optimize its utility in predicting OA and knee replacement outcomes.

This investigation demonstrates an innovative application of qMRI imaging and machine learning methodologies to a large, well-characterized cohort, which reveals structural and compositional biomarkers that are both clinically relevant and interpretable. Our findings show the importance of localized cartilage integrity, meniscal morphology, and bone shape in predicting knee health outcomes, while establishing a foundation for precision-driven disease modeling. Furthermore, the granularity and multi-tissue approach of this framework offer fresh insights into a degenerative disease that imposes substantial economic and quality-of-life impacts on aging populations.[7–10]

This study validates the integration of advanced imaging biomarkers into an extensible pipeline, and establishes a solid foundation for the broader application of digital twin technology in musculoskeletal healthcare. Future research should seek to operationalize the extracted 3D imaging-based representations to validate longitudinal changes in knee health, model treatment responses, and predict disease trajectories. Additionally, efforts to include cohorts that reflect population variability and integrate multimodal data, such as biomechanical signals and kinematic measurements, will be effective for refining this framework for broader clinical applications. By building the infrastructure for a digital twin environment rooted in advanced qMRI imaging and machine learning, we move closer to realizing the potential of precision health strategies for patients with OA.

## Methods

*Dataset*

The dataset analyzed in this study originates from the Osteoarthritis Initiative[33], a longitudinal study conducted across multiple centers. This initiative is designed to investigate the progression of osteoarthritis through a comprehensive imaging dataset collected over a decade, including initial baseline and subsequent visits at 1, 2, 3, 4, 6, and 8 years. The dataset comprises knee radiographs and dual-knee MRIs, capturing the left and right knees of 4,796 participants at each time point. This comprehensive imaging protocol includes 3D double-echo steady-state (3D-DESS) and 2D multi-slice multi-echo (2D-MSME) MRI knee scans. Trained radiologists at the study's central site used the Kellgren-Lawrence grading (KLG) scale,[44] a standard for OA diagnosis, to grade radiographs with high reproducibility. This method was used to assess OA severity, specifically to describe cases with joint space narrowing and osteophyte occurrence. Radiographic OA was defined as a KLG score ≥ 2. Demographic data, including age, sex, BMI, and knee pain scores, were collected at each visit.

The study was conducted in accordance with the Declaration of Helsinki (1964) and its subsequent amendments, as well as other relevant regulations. Institutional review boards (IRBs) at each Osteoarthritis Initiative (OAI) clinical site, including Memorial Hospital of Rhode Island, Ohio State University, University of Pittsburgh, and University of Maryland/Johns Hopkins University, approved the OAI study. The OAI Coordinating Center at the University of California, San Francisco provided IRB approval (approval number 10-00532, Federalwide Assurance #00000068). Additionally, the OAI Clinical Sites Single IRB of Record was study number 2017H0487, Federalwide Assurance #00006378. This study is registered with ClinicalTrials.gov (identifier: NCT00080171). All participants provided written informed consent prior to participation. An independent Observational Study Monitoring Board (OSMB) appointed by the National Institute of Arthritis and Musculoskeletal and Skin Diseases (NIAMS) oversaw the study. The OSMB ensured adherence to ethical research standards and participant safety throughout the study. Further details regarding the demographics and clinical data are provided in Supplementary Data 1 and 2.

*Automatic Knee Joint Tissue Segmentation*

3D-DESS volumes of the OAI dataset underwent automatic bone[45], cartilage[36], and meniscus[25] segmentation using neural net-based models previously trained and validated for the OAI. Model performance was assessed using mean Dice score coefficients, which reflects the average overlap between human-annotated and model-predicted segmentations.

For meniscus and cartilage segmentation, we employed an ensemble of Convolutional Neural Networks (CNNs) consisting of three 3D V-Nets[46] and three 2D U-Net-like[47] models. Each model was independently trained on DESS-we MRIs to segment the menisci and cartilage of the femur, tibia, and patella Our dataset included 176 DESS-we volumes manually annotated by iMorphics.[48] We partitioned these volumes into development and validation sets, reserving 28 volumes for testing. The training emphasized model generalizability, with preprocessing steps including adjustments for orientation, cropping, and signal normalization; the 3D models' training incorporated data augmentation techniques to enhance performance. Training protocols featured Adam optimization[49] and weighted Dice loss, with learning rates adjusted for 2D and 3D models to optimize performance. In the test set, the mean Dice coefficients for tissue segmentation were recorded as follows: meniscus, 0.874 (±0.024), femoral cartilage, 0.890 (±0.023), tibial cartilage, 0.880 (±0.036), and patellar cartilage, 0.850 (±0.068).

To verify the reliability of our automated cartilage thickness measurement system, scan-rescan reliability tests were conducted using 15 sets of MRI scans from the OAI. Compartment-wise analysis revealed that the average absolute variation in measurements was maintained below 0.053 mm (±0.03 mm) for both femoral and tibial compartments. Moreover, the upper 95% confidence limit for the largest average absolute discrepancy between manual and automated measurements was maintained at 0.15 mm. This discrepancy is considerably below the in-plane pixel resolution, highlighting the high precision and reliability of our automated measurement system in accurately quantifying cartilage thickness.

To build bone segmentation models for the femur, tibia, and patella, MRI volumes from 40 OAI participants were selected to represent both healthy individuals and those with OA incidence. A modified 3D V-Net[46] architecture served as our model foundation, which was trained over 185 epochs with specific initializations to ensure effective learning and convergence.
In the analysis of the test dataset, the mean Dice score coefficients along with their 95% confidence intervals were 97.15% (96.56% to 97.74%) for the femur, 97.28% (96.64% to 97.92%) for the tibia, and 95.99% (95.26% to 96.72%) for the patella. In terms of accuracy, the mean point-to-surface distance errors were 0.45 mm (0.23 mm to 0.68 mm), 0.57 mm (0.39 mm to 0.74 mm), and 0.51 mm (0.07 mm to 0.94 mm) for the femur, tibia, and patella, respectively. This corresponds to a precision close to the size of a single voxel.

To further validate our bone segmentation model, we employed an additional set of 60 MRI scans, manually segmented and pseudo-randomly selected from the OAI database. This selection was designed to reflect the demographic diversity and prevalence of osteoarthritis, thereby ensuring the valid applicability of our model across diverse patient profiles.

*Morphometry for Cartilage Thickness*

To assess cartilage thickness from the 3D-DESS MRI data, cartilage thickness maps were derived from the femoral, tibial, and patellar segmentation masks. An Euclidean distance transform method was applied to the morphological skeleton of each of the respective masks. Further details on the automated cartilage thickness measurement can be found in [36].

*Relaxometry for Cartilage $T_2$*

The 2D-MSME image volumes were rigidly aligned with 3D-DESS volumes using the Patient Coordinate System from the DICOM metadata to achieve coregistration of the articular tissue-based imaging biomarkers. The slice resolutions of 2D-MSME were adjusted to match the 3D-DESS resolution using bicubic interpolation. Alignment was applied to all echoes based on the first echo volume, and sagittal slices of 2D-MSME were shifted to match 3D-DESS, creating MSME-DESS registered volumes. The 3D-DESS cartilage segmentation masks were then used to isolate cartilage in MSME-DESS, and cartilage $T_2$ relaxation times were calculated using a three-parameter Levenberg-Marquardt mono-exponential model: $S(TE) = \alpha * \exp(-TSL/T_2) + C$, applied consistently across the three segmentation masks.[28,50]

*Bone and Meniscal Shape*

The segmentation masks of the bone and meniscus were converted from voxel masks to 3D point clouds using the Marching Cubes algorithm[51], followed by rigid registration to a reference point cloud to account for rotational variability. The analysis was conducted individually for the femur, tibia, and patella to accommodate joint positioning variations. The surfaces of these bones were registered with Iterative Closest Point matching to retain distinct shape features for a Statistical Shape Model (SSM). A reference knee was iteratively defined using registration error metrics, and its landmarks were mapped onto all dataset surfaces using an automatic landmark-matching algorithm[52] based on local curvatures. This comprehensive approach enables downstream statistical analysis using the geometric features of the surface topology.

*Bone surface projection of Cartilage Biomarkers*

Utilizing triangulated meshes generated from cartilage masks via the Marching Cubes algorithm,[51] cartilage thickness and $T_2$ values were projected onto the articular bone surface. Thickness values were aimed at perpendicular points, and superficial, deep, and total average $T_2$ values for specific cartilage sections were similarly mapped based on the intersection of normal vectors, facilitating the averaging of these measures across the cartilage cross-section.

*Dimensionality Reduction and Statistical Shape Modeling*

We applied Principal Component Analysis (PCA) to each tissue biomarker independently and systematically. This approach aimed to reduce data dimensionality while preserving geometric variations that pertain to each tissue and biomarker. Data preparation involved careful identification of tissue-specific landmarks; these points defined the anatomical geometry under consideration. Each surface in the OAI dataset was represented by 3D coordinates assigned to these landmarks[53], effectively transforming the surface into a single point in a 3L-dimensional space, where 'L' denotes the number of landmarks. We then computed the mean surface for each tissue-biomarker combination by averaging all surfaces. This reference surface enabled precise discernment and interpretation of shape variations within the dataset.

Our dimensionality reduction process distilled the dataset into 110 principal components (PCs), arranged according to defined tissue types and imaging biomarkers. The PCs spanned three tissue types (Femur,

Patella, and Tibia) as well as three imaging biomarkers (Bone Shape, Cartilage Thickness, and Cartilage $T_2$ Relaxation Time). For each tissue-biomarker pairing, we extracted 10 PC modes, capturing features that included Femur Bone Shape, Femur Cartilage Thickness, and Femur Cartilage $T_2$, among others. We also incorporated the Medial and Lateral Meniscus as separate tissues with Meniscus Shape as their imaging biomarker, adding another 20 PC modes. This classification yielded a total of 110 PC mode features, offering a broad perspective on dataset variability related to OA (Supplementary Data 3). Supplementary Figure 1 further details how these PC modes retain cumulative geometric variance and maintain the dataset's compactness.

*PC Mode Visualization*

Synthetic images were generated to interpret PC modes, which capture unique aspects of the imaging biomarker features, each independent due to PCA's orthonormal basis. The effect of each mode was observed by generating a new 3D surface and comparing its difference from the average surface for a given feature. The physical representation of the modes was further explored by varying each mode's value within a range of mean ± 3 standard deviations (SD). This variation allowed for the observation of landmark displacements from the average surface, visualized using three-dimensional colored meshes[28].

*PCA Mode Interpretation*

Radiologist experts collectively analyzed each mode within the 110-dimensional PCA space using a customized MATLAB-based software application (Figure 2). This interface, designed to be accessible, assists with choosing and visualizing tissue types, biomarkers, and PCA modes, and allows interactive engagement with the underlying data. Users can manipulate and save visual representations of PCA modes, as well as input their interpretations alongside displayed community consensus. A navigable, worksheet-like tab supports the tracking of user analysis and interpretation efforts; this ensures an organized path toward a collective viewpoint. The application's flexible architecture can accommodate future additions; specifically, it may incorporate new knee image feature embeddings into the visual space as more data becomes available. The PCA interpretation application's source code is provided at https://github.com/gabbieHoyer/OAI-PC-mode-interpreter, granting the research community broad access to its functionalities.

*Cohort Construction and Matching for Outcome Analysis*

In our examination of the 110-dimension qMRI feature space, cohort matching analysis proved instrumental in revealing complex biological signatures and their potential causal influence on knee-related patient outcomes. We employed statistical cohort matching and analysis in two distinct sets. This approach minimized confounding influences from demographic factors, physical activity levels, and pain, allowing for a more targeted assessment of individual knee imaging biomarkers and their effects on the outcomes of interest.

Specifically, cohort matching was conducted for a) Osteoarthritis (OA) Incidence and b) Knee Replacement (partial/total). The OA Incidence cohort comprised control knees without radiographic OA (Kellgren-Lawrence grade [KLG] < 2) at all time points (baseline to 96 months) and the OA Incidence group, which included knees initially without OA but with subsequent development of OA (KLG ≥ 2 at any follow-up). For the Knee Replacement analysis, the control group consisted of knees without knee replacement surgery, while the outcome group included those undergoing partial or total knee replacement after baseline evaluation.

Each cohort was carefully constructed to follow the outlined criteria without deviation. The demographic and clinical factors used in this analysis are detailed in the 'Demographic and Clinical Factor Definitions' table (Supplementary Data 1). These factors were initially operationalized from the Osteoarthritis Initiative (OAI)[33] and the Knee Osteoarthritis New Approaches (KNOAP) Challenge.[54] They include a broad range of demographic data and clinical measures such as knee pain and quality assessments, offering the contextual foundation needed for our dataset and all subsequent analyses.

*Data Preprocessing and Sensitivity Analysis*

In the data preprocessing phase, particular attention was given to ensuring the integrity and consistency of our dataset, which is foundational for the reliability of our analyses. We focused on the baseline imaging biomarker PC modes of the OAI subjects, along with demographic and clinical factors. To streamline our analytical approach and produce clearer, more interpretable results, we exclusively analyzed MRI data from the right knee, a choice consistent with existing literature.[50] Outcomes for OA Incidence and Knee Replacement were identified using binary variables. Partial and total knee replacements were grouped into a single outcome category, contrasted against a control group without surgery.

We addressed data completeness by excluding subjects who lacked imaging values for any PC mode features, thereby maintaining the quality of our dataset. Additionally, we evaluated the dataset at the feature level, excluding any demographic or clinical factors with more than 5% missing data across subjects to ensure that only reliable data informed our analysis. For the remaining minimal data gaps, multiple imputation was performed for both Control and Knee Outcome cohorts using the sklearn[55] "IterativeImputer." This tool applied a Random Forest Classifier for categorical variables and a Random Forest Regressor for continuous variables, chosen for their effectiveness in handling missing data—using mode value for categorical and median value for continuous variables.

To validate the efficacy of our imputation approach, we conducted a sensitivity analysis using Kolmogorov-Smirnov tests to compare the distributions of covariates before and after imputation; this evaluation confirmed the preservation of the data's overall structure. The findings of this assessment, which demonstrate the reliability of our data preparation procedures, appear in Supplementary Data 4 and 5. After resolving missing values, we transformed categorical variables into numeric form and standardized continuous variables with sklearn's[55] "StandardScaler." Applying this scaling step ensured that all features shared a consistent range, thus promoting accuracy and reliability in subsequent analyses.

*Cohort Matching*

In our study, cohort matching was a critical step to ensure a balanced comparison between Control and Knee Outcome groups, considering a wide array of demographic and clinical factor covariates. To achieve this, we employed the t-Distributed Stochastic Neighbor Embedding (t-SNE) technique[35], known for its effectiveness in high-dimensional data visualization and clustering, as a robust mechanism for cohort matching. Specifically, after standardizing the data, it was embedded into a 3D t-SNE space, with the perplexity set to $\sqrt{N}$ samples to optimize the balance between local and global data structures. We then calculated the Euclidean distance between each data point in the embedded space and sorted the resultant distance matrix. Cohort matches were identified by selecting pairs with the shortest Euclidean distance, effectively linking a subject from the Knee Outcome cohort with a Control subject with the most similar clinical profile. Given the imbalance in cohort sizes, matching was performed with replacement, meaning some control subjects were matched to multiple knee outcome subjects to ensure reliable comparisons while maintaining the integrity of the dataset. A flowchart detailing the subject selection process for our Cohort Matching Analyses is provided in Figure 3.

The OA Incidence Clinical Cohort Analysis was conducted on $N_{Clinical}$ = 2,378 participants, with $n_{Control,Clinical}$ = 319 subjects in the Control cohort and $n_{OA\text{-}Incidence,Clinical}$ = 357 in the OA Incidence cohort. Additionally, the Knee Replacement Clinical Cohort Analysis was conducted on $N_{Clinical}$ = 4,283 participants, with $n_{Control,Clinical}$ = 233 subjects in the Control cohort and $n_{KR,Clinical}$ = 253 in the Knee Replacement cohort. Information about the demographic and clinical characteristics of the participants within these Cohort Analyses can be found in Supplementary Data 6 and 7.

*Evaluation of Matching Technique*

To determine how effective our cohort-matching approach was, we initially calculated general descriptive statistics for each covariate before and after the matching procedure; this approach offered a clear overview of cohort characteristics and confirmed that the Control and Knee Outcome groups were initially comparable. The covariates evaluated included continuous measures such as age, BMI, and pain scores, as well as categorical variables such as gender, joint tenderness, and injury history.

We assessed the effect size and association strength in matched groups for each covariate using Point Biserial Correlation Coefficients and Cohen's d, also referred to as the Standardized Mean Difference (SMD), for continuous variables; Cohen's d was calculated as the mean difference divided by the combined standard deviation. These measures were selected for their ability to effectively quantify the magnitude of differences and associations. For categorical covariates, we used Cramer's V to determine the strength of association between the groups.

As a preliminary step, we assessed the normality and homoscedasticity of continuous covariates in the matched cohorts using Shapiro-Wilk and Anderson-Darling tests, which provided W statistics, p-values, and significance levels to identify variables that violated normality assumptions. These results guided the selection of non-parametric methods for subsequent analyses. For null hypothesis testing, the Wilcoxon Rank Sum Test (two-sided) was utilized for continuous covariates, while the Chi-squared Test was applied to categorical covariates. No corrections for multiple comparisons were applied, as these analyses served as preliminary evaluations of matched cohort quality. Both tests were conducive in determining whether significant differences existed between the cohorts post-matching.

According to the outcomes recorded in Supplementary Data 8-11, our cohort-matching technique demonstrates reliable quality. As a result, we can assert that the matching process strengthened both the integrity and credibility of the analyses.

*Cross-Sectional Statistical Analyses*

A focused statistical exploration was performed for the Clinical Cohort Analysis, wherein tissue-biomarker PC features were examined for their significance relative to knee outcomes. Our null hypothesis (H0) asserted no differential expression of tissue-biomarker PC features between the Control and Knee Outcome cohorts, while the alternative hypothesis (HA) posited a significant differential expression.

Before proceeding to hypothesis testing for the PC features, Shapiro-Wilk and Anderson-Darling tests were performed at a 5% significance level to assess the assumption of normality for all 110 PC modes. These tests determined the appropriateness of statistical tests for each variable. Shapiro-Wilk tests reported W statistics and p-values, while Anderson-Darling tests provided test statistics and critical significance levels. Results for these normality tests can be found in Supplementary Data 12 and 13.

For the analysis of continuous variables, we employed two-sided Wilcoxon Rank Sum tests, as dictated by the results of the preliminary normality tests, to identify significant differences between the medians of two related groups. These tests were performed at an $\alpha=0.05$ significance level, with degrees of freedom (df = 356) for the OA Incidence cohort and df=252 for the Knee Replacement cohort. Confidence intervals (95% and 99%) were calculated via bootstrapping (N=1000 iterations) for each of the 110 PC mode variables. This approach allowed a thorough exploration of the data.

To manage the Type I error rate across multiple comparisons of the 110 PC mode variables, we utilized the Benjamini-Hochberg correction, adjusting p-values using $P'(i) = min(\ m\ i \times P(i), P'(i+1)\ )$, where m as the total number of tests (110 in this case), and $P'(i+1)$ as the next highest adjusted p-value. This method was used for the main results in the paper, balancing control of false positives while maintaining statistical power. Detailed results of the statistical tests, including correction-adjusted p-values, are presented in the supplementary materials (Supplementary Data 14 and 15). This thorough statistical approach provided a detailed evaluation of tissue-biomarker PC features in relation to knee outcomes.

*Multivariate Regression Analyses*

We conducted multivariate regression analyses in our study to better understand the contributions of the imaging biomarker features to knee-related outcomes, specifically OA incidence and knee replacement. Our methodology utilized the matched pairs from our cohort groups, incorporating 110 principal components alongside key covariates like gender, age, BMI, and pain scores. Our comprehensive multivariate approach integrated elastic net regularization with a generalized linear model (GLM), employing a binomial distribution with a logit-link function and iterating over 1,000 bootstrap samples; statistical modeling was performed using the Statsmodels[56] library in Python. The GLM analysis was two-sided.

To identify features consistently contributing to the model's predictive power, we employed a variant of stability selection.[57] For each sampling, feature coefficients (Supplementary Data 16 and 19) with an absolute value exceeding a threshold of 1e-5 were deemed relevant, and their selection occurrence was recorded. This process produced values for the relative frequency of feature selection and absolute mean coefficients, which represent the average effect size of each feature. From the product of this relative frequency and absolute mean coefficient, a 'weighted importance' score was derived, which guided consistent identification of significant features (Supplementary Data 17 and 20). Features were retained if their weighted importance fell within the upper quartile range. Restricting the selection to features in the upper quartile range meant that the final analysis would focus solely on those with consistently high stability in bootstrapped samples and substantial effect sizes. This measure decreased the likelihood of choosing features driven by spurious associations. As a result, we emphasized features offering strong and consistent predictive contributions, thereby reinforcing their value as dependable biomarkers for knee-related outcomes.

The refined feature set was then applied to a final logistic regression, where we evaluated the impact using standard model parameters, including coefficients, standard errors, p-values (significance level set at $p < 0.05$), 95% and 99% confidence intervals (CI), and odds ratios (Supplementary Data 18 and 21). Odds ratios provided interpretable effect sizes, indicating the multiplicative change in the odds of knee-related outcomes associated with one-unit changes in predictors. Additionally, model efficacy was assessed using Pseudo R-squared, Pearson Chi-square metrics, and other key metrics (Supplementary Data 22 and 23). This evaluation confirmed that our approach reliably identified biomarkers associated with knee-related health outcomes.

## Data availability

All data supporting the findings of this study are available within the manuscript and its Supplementary Information. The primary data table developed for this study, detailing the OAI Imaging Biomarkers 100-dimensional PCA Feature Space for the Baseline Timepoint, is provided as Supplementary Data 3. Additional data sources include the publicly available Osteoarthritis Initiative (OAI[33]) variables (https://nda.nih.gov/oai), KNOAP[54] variables (https://knoap2020.grand-challenge.org/), and the OAI Tissue Segmentations[25] dataset available at Kaggle (https://www.kaggle.com/datasets/kgaooo/oai-tissue-segmentations).

The publicly available datasets used in this study (OAI and KNOAP) are hosted on their respective platforms and can be accessed under their terms of use. The OAI Tissue Segmentations dataset on Kaggle provides segmentations generated by co-authors, supporting replication and validation of the findings. Researchers interested in accessing these datasets should refer to the links provided above.

## Code availability

Custom code used in this study is available in publicly accessible repositories. The code for the 110-dimensional imaging biomarker principal component visualization tool, which includes high-resolution images and expertly defined interpretations, can be accessed at https://github.com/gabbieHoyer/OAI-PC-mode-interpreter. The code for all data processing, dimensionality reduction, cohort matching, and statistical analysis is hosted at https://github.com/gabbieHoyer/OAI_digital_twins. Additionally, the pipeline for meniscus shape modeling is available at https://github.com/kenneth-gao/meniscus_ssm/tree/master. These repositories are openly accessible and provide all necessary information to replicate and extend the results presented in this study.

## Acknowledgements

This work is supported by the National Institute of Arthritis and Skin Diseases (NIH/NIAMS R33AR073552/R00AR070902). The OAI, a public-private partnership supported by NIH contracts (N01-AR-2-2258; N01-AR-2-2259; N01-AR-2-2260; N01-AR-2-2261; N01-AR-2-2262), also received funding through the Foundation for the NIH, with contributions from Merck, Novartis, GlaxoSmithKline, and Pfizer.

## Authors' contributions

All authors contributed to the conception and design of the study. All authors reviewed and approved the final manuscript. G.H. served as the primary analyst and coordinated the analytical framework, conducted data and statistical analyses, and introduced a new analytical tool. G.H. prepared the initial manuscript draft and designed the figures. F.G. and J.L. interpreted PCA modes and conducted in-depth analyses of 110 features related to knee joint anatomy and osteoarthritis, a contribution that enriched the study's understanding of OA imaging biomarkers. F.J. contributed to the statistical analysis framework and guided the refinement of advanced analytical methods. K.G. assisted with data acquisition and processing during the early stages and influenced the overall study design. V.P. and S.M. conceived the study, secured funding, and provided valuable input during manuscript revisions.

## Competing interests

All authors declare no financial or non-financial competing interests.

## References

1. Deshpande, B. R. *et al.* Number of persons with symptomatic knee osteoarthritis in the US: impact of race and ethnicity, age, sex, and obesity. *Arthritis care & research* **68**, 1743–1750 (2016).
2. Jacobs, C. A. *et al.* Development of a mind body program for obese knee osteoarthritis patients with comorbid depression. *Contemporary Clinical Trials Communications* **21**, 100720 (2021).
3. Cisternas, M. G. *et al.* Alternative methods for defining osteoarthritis and the impact on estimating prevalence in a US population-based survey. *Arthritis care & research* **68**, 574–580 (2016).

4. Zhang, Y. & Jordan, J. M. Epidemiology of osteoarthritis. *Clinics in geriatric medicine* **26**, 355–369 (2010).
5. Ackerman, I. N. *et al.* The substantial personal burden experienced by younger people with hip or knee osteoarthritis. *Osteoarthritis and Cartilage* **23**, 1276–1284 (2015).
6. Losina, E., Thornhill, T. S., Rome, B. N., Wright, J. & Katz, J. N. The dramatic increase in total knee replacement utilization rates in the United States cannot be fully explained by growth in population size and the obesity epidemic. *JBJS* **94**, 201–207 (2012).
7. Xu, Y. & Wu, Q. Trends and disparities in osteoarthritis prevalence among US adults, 2005–2018. *Scientific Reports* **11**, 21845 (2021).
8. Barbour, K. E. Vital signs: prevalence of doctor-diagnosed arthritis and arthritis-attributable activity limitation—United States, 2013–2015. *MMWR. Morbidity and mortality weekly report* **66**, (2017).
9. Helmick, C. G. *et al.* Estimates of the prevalence of arthritis and other rheumatic conditions in the United States: Part I. *Arthritis & Rheumatism* **58**, 15–25 (2008).
10. Hunter, D. J. & Bierma-Zeinstra, S. Osteoarthritis. *The Lancet* **393**, 1745–1759 (2019).
11. Eckstein, F., Burstein, D. & Link, T. M. Quantitative MRI of cartilage and bone: degenerative changes in osteoarthritis. *NMR in Biomedicine: An International Journal Devoted to the Development and Application of Magnetic Resonance In vivo* **19**, 822–854 (2006).
12. Chevalier, X., Eymard, F. & Richette, P. Biologic agents in osteoarthritis: hopes and disappointments. *Nat Rev Rheumatol* **9**, 400–410 (2013).
13. Ghouri, A. & Conaghan, P. G. Update on novel pharmacological therapies for osteoarthritis. *Therapeutic Advances in Musculoskeletal* **11**; 10.1177/1759720X19864492 (2019).
14. RODRIGUEZ-MERCHAN, E. C. The current role of disease-modifying osteoarthritis drugs. *ABJS* (2022).
15. Bhosale, A. M. & Richardson, J. B. Articular cartilage: structure, injuries and review of management. *British medical bulletin* **87**, 77–95 (2008).
16. Felson, D. T. *et al.* Osteoarthritis: new insights. Part 2: treatment approaches. *Annals of internal medicine* **133**, 726–737 (2000).
17. Li, X. & Majumdar, S. Quantitative MRI of articular cartilage and its clinical applications. *Magnetic Resonance Imaging* **38**, 991–1008 (2013).
18. Wirth, W. *et al.* Predictive and concurrent validity of cartilage thickness change as a marker of knee osteoarthritis progression: data from the Osteoarthritis Initiative. *Osteoarthritis and cartilage* **25**, 2063–2071 (2017).

19. Eckstein, F. *et al.* Brief report: cartilage thickness change as an imaging biomarker of knee osteoarthritis progression: data from the Foundation for the National Institutes of Health Osteoarthritis Biomarkers Consortium. *Arthritis & rheumatology* **67**, 3184–3189 (2015).
20. Bredbenner, T. L. *et al.* Statistical shape modeling describes variation in tibia and femur surface geometry between Control and Incidence groups from the osteoarthritis initiative database. *Journal of biomechanics* **43**, 1780–1786 (2010).
21. Neogi, T. *et al.* Magnetic resonance imaging–based three-dimensional bone shape of the knee predicts onset of knee osteoarthritis: data from the Osteoarthritis Initiative. *Arthritis & Rheumatism* **65**, 2048–2058 (2013).
22. Hunter, D. *et al.* Longitudinal validation of periarticular bone area and 3D shape as biomarkers for knee OA progression? Data from the FNIH OA Biomarkers Consortium. *Annals of the rheumatic diseases* **75**, 1607–1614 (2016).
23. Berthiaume, M. J. *et al.* Meniscal tear and extrusion are strongly associated with progression of symptomatic knee osteoarthritis as assessed by quantitative magnetic resonance imaging. *Annals of the rheumatic diseases* **64**, 556–563 (2005).
24. Dube, B. *et al.* Where does meniscal damage progress most rapidly? An analysis using three-dimensional shape models on data from the Osteoarthritis Initiative. *Osteoarthritis and Cartilage* **26**, 62–71 (2018).
25. Gao, K. T. *et al.* Large-Scale Analysis of Meniscus Morphology as Risk Factor for Knee Osteoarthritis. *Arthritis & Rheumatology* **75**, 1958–1968 (2023).
26. Robinson, W. H. *et al.* Low-grade inflammation as a key mediator of the pathogenesis of osteoarthritis. *Nature Reviews Rheumatology* **12**, 580–592 (2016).
27. Loeser, R. F., Collins, J. A. & Diekman, B. O. Ageing and the pathogenesis of osteoarthritis. *Nature Reviews Rheumatology* **12**, 412–420 (2016).
28. Pedoia, V., Lee, J., Norman, B., Link, T. M. & Majumdar, S. Diagnosing osteoarthritis from T2 maps using deep learning: an analysis of the entire Osteoarthritis Initiative baseline cohort. *Osteoarthritis and cartilage* **27**, 1002–1010 (2019).
29. Hunter, D. J., Nevitt, M., Losina, E. & Kraus, V. Biomarkers for osteoarthritis: current position and steps towards further validation. *Best practice & research Clinical rheumatology* **28**, 61–71 (2014).
30. Committee on Foundational Research Gaps and Future Directions for Digital Twins *et al. Foundational Research Gaps and Future Directions for Digital Twins*. 26894 (National Academies Press, Washington, D.C., 2024).
31. *Digital Twin: Definition & Value. An AIAA and AIA Position Paper*. (2020).

32. *White Paper: The Value of Digital Twin Technology*. https://www.siemens-healthineers.com/en-us/services/value-partnerships/asset-center/white-papers-articles/value-of-digital-twin-technology (2019).
33. Osteoarthritis Initiative (OAI), a multi-center, longitudinal, prospective observational study of knee osteoarthritis, sponsored by the National Institutes of Health (NIH) and private industry partners.
34. Roos, E. M. & Lohmander, L. S. The Knee injury and Osteoarthritis Outcome Score (KOOS): from joint injury to osteoarthritis. *Health Qual Life Outcomes* **1**, 64 (2003).
35. Maaten, L. & Hinton, G. Visualizing data using t-SNE. *Journal of machine learning research* **9**, (2008).
36. Iriondo, C. *et al.* Towards understanding mechanistic subgroups of osteoarthritis: 8-year cartilage thickness trajectory analysis. *Journal of Orthopaedic Research®* **39**, 1305–1317 (2021).
37. Baum, T. *et al.* Changes in knee cartilage T2 values over 24 months in subjects with and without risk factors for knee osteoarthritis and their association with focal knee lesions at baseline: Data from the osteoarthritis initiative. *Magnetic Resonance Imaging* **35**, 370–378 (2012).
38. Kretzschmar, M. *et al.* Spatial distribution and temporal progression of T2 relaxation time values in knee cartilage prior to the onset of cartilage lesions – data from the Osteoarthritis Initiative (OAI). *Osteoarthritis and Cartilage* **27**, 737–745 (2019).
39. Kawahara, T., Sasho, T. & Katsuragi, J. Relationship between knee osteoarthritis and meniscal shape in observation of Japanese patients by using magnetic resonance imaging. *J Orthop Surg* **12**, 97 (2017).
40. Eckstein, F., Wirth, W., Lohmander, L. S., Hudelmaier, M. I. & Frobell, R. B. Five-year followup of knee joint cartilage thickness changes after acute rupture of the anterior cruciate ligament. *Arthritis & rheumatology* **67**, 152–161 (2015).
41. Barr, A. J. *et al.* The relationship between three-dimensional knee MRI bone shape and total knee replacement—a case control study: data from the Osteoarthritis Initiative. *Rheumatology* **55**, 1585–1593 (2016).
42. Bowes, M. A., Vincent, G. R., Wolstenholme, C. B. & Conaghan, P. G. A novel method for bone area measurement provides new insights into osteoarthritis and its progression. *Annals of the rheumatic diseases* **74**, 519–525 (2015).
43. Sun, T. *et al.* The Digital Twin: A Potential Solution for the Personalized Diagnosis and Treatment of Musculoskeletal System Diseases. *Bioengineering* **10**, 627 (2023).
44. Kellgren, J. H. & Lawrence, J. S. Radiological assessment of osteo-arthrosis. *Ann Rheum Dis* **16**, 494–502 (1957).

45. Morales Martinez, A. *et al.* Learning osteoarthritis imaging biomarkers from bone surface spherical encoding. *Magnetic resonance in medicine* **84**, 2190–2203 (2020).
46. Milletari, F., Navab, N. & Ahmadi, S.-A. V-Net: Fully Convolutional Neural Networks for Volumetric Medical Image Segmentation. Preprint at https://doi.org/10.48550/ARXIV.1606.04797 (2016).
47. Ronneberger, O., Fischer, P. & Brox, T. U-Net: Convolutional Networks for Biomedical Image Segmentation. in *Medical Image Computing and Computer-Assisted Intervention – MICCAI 2015* (eds. Navab, N., Hornegger, J., Wells, W. M. & Frangi, A. F.) vol. 9351 234–241 (Springer International Publishing, Cham, 2015).
48. *White Paper: Imorphics OA Knee MRI Measurements*. www.imorphics.com (2017).
49. Kingma, D. P. & Ba, J. Adam: A Method for Stochastic Optimization. Preprint at https://doi.org/10.48550/ARXIV.1412.6980 (2014).
50. Morales, A. G. *et al.* Uncovering associations between data-driven learned qMRI biomarkers and chronic pain. *Scientific reports* **11**, 21989 (2021).
51. Lorensen, W. E. & Cline, H. E. Marching cubes: A high resolution 3D surface construction algorithm. in *Seminal graphics: pioneering efforts that shaped the field* 347–353 (1998).
52. Lombaert, H., Grady, L., Polimeni, J. R. & Cheriet, F. FOCUSR: feature oriented correspondence using spectral regularization–a method for precise surface matching. *IEEE transactions on pattern analysis and machine intelligence* **35**, 2143–2160 (2012).
53. Pedoia, V. *et al.* Three-dimensional MRI-based statistical shape model and application to a cohort of knees with acute ACL injury. *Osteoarthritis and cartilage* **23**, 1695–1703 (2015).
54. Hirvasniemi, J. *et al.* The KNee OsteoArthritis Prediction (KNOAP2020) challenge: An image analysis challenge to predict incident symptomatic radiographic knee osteoarthritis from MRI and X-ray images. *Osteoarthritis and cartilage* **31**, 115–125 (2023).
55. Pedregosa, F. *et al.* Scikit-learn: Machine learning in Python. *the Journal of machine Learning research* **12**, 2825–2830 (2011).
56. Perktold, J. *et al.* jbrockmendel, grana6. doi:10.5281/zenodo.10378921.
57. Meinshausen, N. & Bühlmann, P. Stability selection. *Journal of the Royal Statistical Society: Series B (Statistical Methodology* **72**, 417–473 (2010).